\documentclass{iau} 
\usepackage{graphicx}

\title[Gravity waves in massive stars] 
{Massive star interiors revealed by \\ gravity wave asteroseismology and high-resolution spectroscopy}

\author[Dominic M. Bowman]   
{Dominic M. Bowman$^{1}$}
\affiliation{
$^1$Institute of Astronomy, KU Leuven, Celestijnenlaan 200D, 3001 Leuven, Belgium \\ email: {\tt dominic.bowman@kuleuven.be}
}

\pubyear{2022}
\volume{361}  
\jname{Massive Stars Near and Far}
\editors{N. St-Louis, J. S. Vink \& J. Mackey, eds.}
\begin{document}

\maketitle

\begin{abstract}
In recent years, it has been discovered that massive stars commonly exhibit a non-coherent form of variability in their light curves referred to as stochastic low frequency (SLF) variability. Various physical mechanisms can produce SLF variability in such stars, including stochastic gravity waves excited at the interface of convective and radiative regions, dynamic turbulence generated in the near-surface layers, and clumpy winds. Gravity waves in particular are a promising candidate for explaining SLF variability as they can be ubiquitously generated in main sequence stars owing to the presence of a convective core, and because they provide the large-scale predominantly tangential velocity field required to explain macroturbulence in spectral line fitting. Here, I provide an overview of the methods and results of studying SLF variability in massive stars from time series photometry and spectroscopy.

\begin{keywords}
stars: early-type, stars: oscillations (including pulsations), stars: interiors, waves.
\end{keywords}
\end{abstract}

\firstsection 
\section{Introduction}

Modern time-series space photometry, such as from the CoRoT \cite[(Auvergne et al. 2009)]{Auvergne2009}, Kepler/K2 \cite[(Borucki et al. 2010; Howell et al. 2014)]{Borucki2010, Howell2014} and TESS \cite[(Ricker et al. 2015)]{Ricker2015} missions, have revealed in the last decade that massive stars show a diverse range of variability mechanisms \cite[(see Bowman 2020 for a review)]{Bowman2020c}. In particular, in addition to a high variability fraction above 90\% for early-type star light curves, which includes coherent pulsation modes, binarity and rotational modulation (see e.g. \cite[Burssens et al. 2020]{Burssens2020}), massive stars almost ubiquitously show stochastic low-frequency (SLF) variability \cite[(Bowman et al. 2019a, 2019b, 2020)]{Bowman2019a, Bowman2019b, Bowman2020b}. SLF variability is also common across the different evolutionary stages of massive stars including blue and yellow supergiants, Wolf Rayet stars and luminous blue variables (see e.g., \cite[Bowman et al. 2019b; Dorn-Wallenstein et al. 2019; Naz{\'e} et al. 2021; Elliott et al. 2022, Lenoir-Craig et al. 2022]{Bowman2019b, Dorn-Wallenstein2020, Naze2021, Elliott2022, Lenoir-Craig2022}), which show similar amplitude spectra to those of main sequence stars. There are currently a few proposed mechanisms to explain SLF variability in massive stars, but it is unclear which may dominate in particular parameter regimes. Hence the investigation of SLF variability in space photometry has allowed for a novel method to understanding the interiors of massive stars \cite[(Bowman 2020)]{Bowman2020c}.

In their original pilot study, \cite[Bowman et al. (2019a)]{Bowman2019a} analysed 35 early-type stars from the CoRoT mission to demonstrate how solar granulation scaling laws cannot explain SLF variability in massive stars. Using a much larger sample of over 160 massive stars from the Kepler/K2 and TESS missions, including galactic and Large Magellanic Cloud (LMC) stars, \cite[Bowman et al. (2019b)]{Bowman2019b} demonstrated that essentially all massive stars exhibit SLF variability, and that its morphology is dependent on the brightness of the star. More recently, \cite[Bowman et al. (2020)]{Bowman2020b} studied a sample of galactic O and B stars from the TESS mission and combined it with high resolution spectroscopy \cite[(Sim{\'o}n-D{\'{\i}}az et al. 2011; Burssens et al. 2020)]{Simon-Diaz2011d, Burssens2020a} to show how the morphology of SLF variability was correlated with mass, age, and amount of macroturbulence needed to fit the spectral line broadening in the massive star regime. 

In an approach similar to the forward asteroseismic modelling of coherent pulsation modes (see \cite[Aerts 2021]{Aerts2021}), the exploitation of SLF variability in time series photometry offers a route to improving our understanding of stellar structure and evolution \cite[(Bowman 2020)]{Bowman2020c}. In these proceedings, I review the current state-of-the-art in the methods and analysis of SLF variability in massive stars.


\section{Stochastic Low Frequency Variability}

The SLF variability in massive stars typically covers a broad period (and frequency) range from several days up to minutes. The amplitudes, on the other hand, range between tens of $\mu$mag up to tens of mmag in broadband time series photometry \cite[(Bowman et al. 2019b)]{Bowman2019b}. Of course, these metrics depend on the length of the light curve, its cadence, wavelength passband and photometric precision. For example, the wavelength passband of the TESS mission is redder to those of the Kepler/K2 and CoRoT missions, which means TESS is slightly less sensitive to the changes in brightness of blue stars \cite[(Bowman et al. 2019a)]{Bowman2019a}. Similarly, with higher photometric precision and longer duration light curves, smaller changes in flux in a light curve can be measured and an improved frequency resolution is achieved in an amplitude spectrum, respectively. Lastly, a shorter cadence yields a higher Nyquist frequency, which is necessary to unambiguously probe short-period variability in massive stars and avoid instrumental suppression \cite[(see Bowman 2017)]{Bowman2017}. Nevertheless, space mission light curves show ubiquitous SLF variability in massive stars when high enough quality time series data are used \cite[(Bowman et al. 2019b)]{Bowman2019b}.

\begin{figure}[b]
\begin{center}
\includegraphics[width=0.49\textwidth]{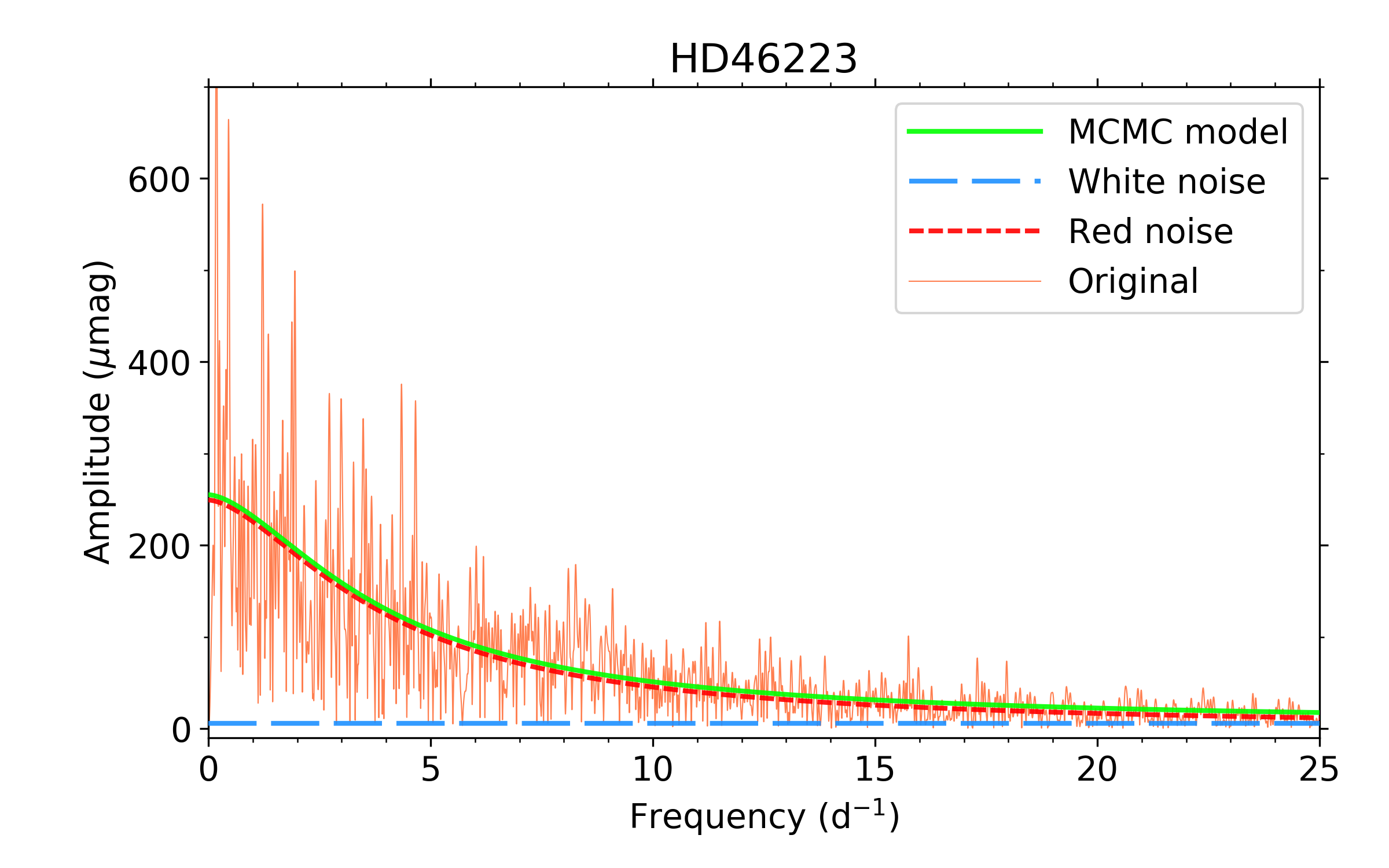} 
\includegraphics[width=0.49\textwidth]{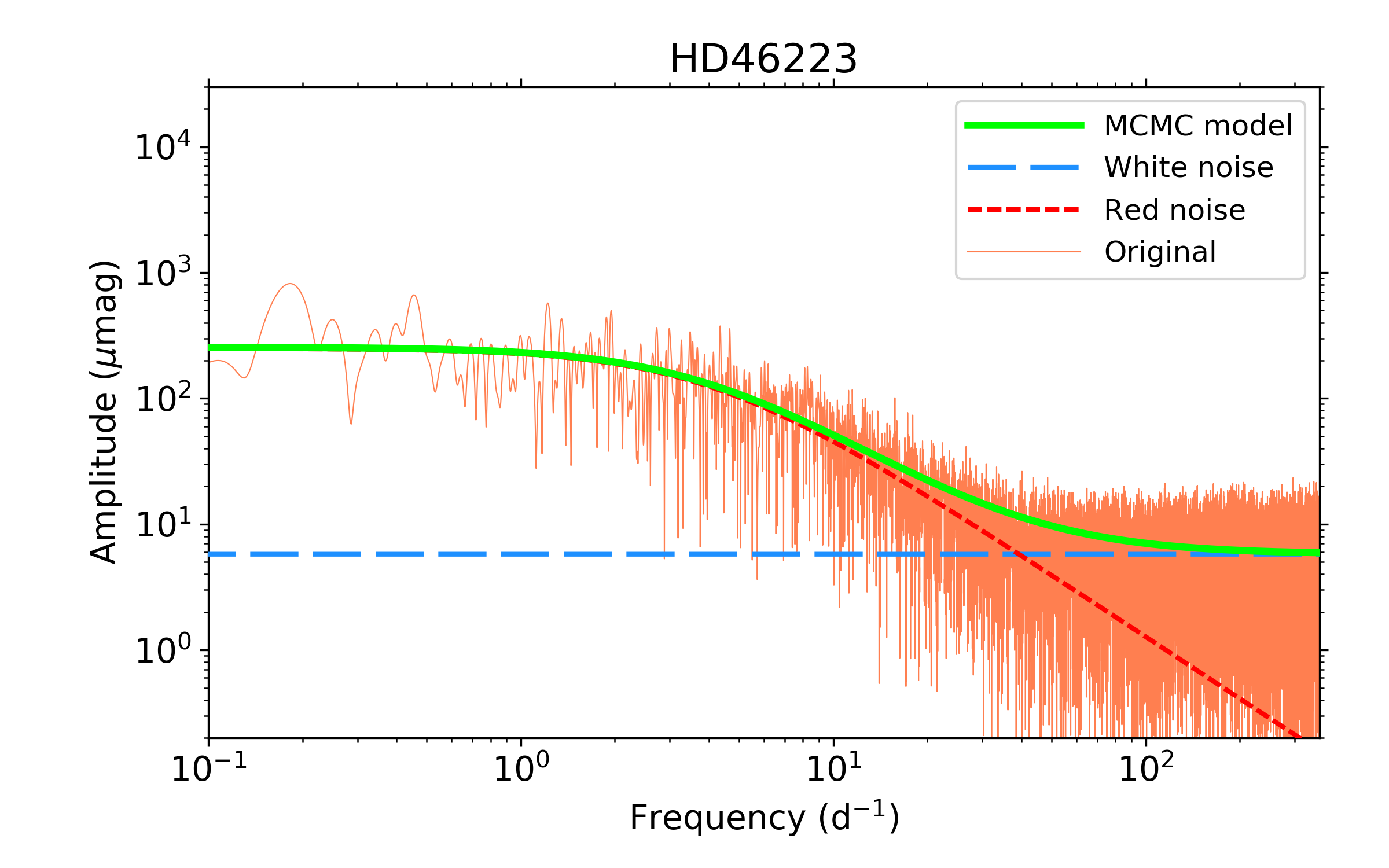} 
\caption{Amplitude spectrum of the 25-d sector 6 TESS light curve of the O4V((f)) star HD~46223. Left and right panels are the same data plotted in a linear-linear and log-log format, respectively. Figures adapted from \cite[Bowman et al. (2020)]{Bowman2020b}, their figure 1.}
\label{figure: FT}
\end{center}
\end{figure}

To characterise the morphology of SLF variability in the Fourier domain, the semi-Lorentzian function:

\begin{equation}
\alpha \left( \nu \right) = \frac{ \alpha_{0} } { 1 + \left( \frac{\nu}{\nu_{\rm char}} \right)^{\gamma}} + C_{\rm w} ~ ,	
\label{equation: red noise}
\end{equation}

\noindent has been used to quantify the characteristic frequency, $\nu_{\rm char}$, characteristic amplitude, $\alpha_0$ and steepness, $\gamma$, of the profile, with $C_{\rm W}$ being a white noise term. An example of observed SLF variability in the amplitude spectrum of a massive star, using the TESS data of the O4V((f)) star HD~46223, is shown in Fig.~\ref{figure: FT}. In the left and right panels, the linear and logarithmic amplitude spectra are shown, each with the best-fit profile using Eqn.~(\ref{equation: red noise}) as determined by \cite[Bowman et al. (2020)]{Bowman2020b}. HD~46223 was also observed by the CoRoT mission and the characteristic amplitude of its SLF variability was somewhat smaller in the TESS data, as expected \cite[(Bowman et al. 2019a)]{Bowman2019a}.


\section{Trends in Mass and Age}

\begin{figure}[b]
\begin{center}
\includegraphics[width=0.49\textwidth]{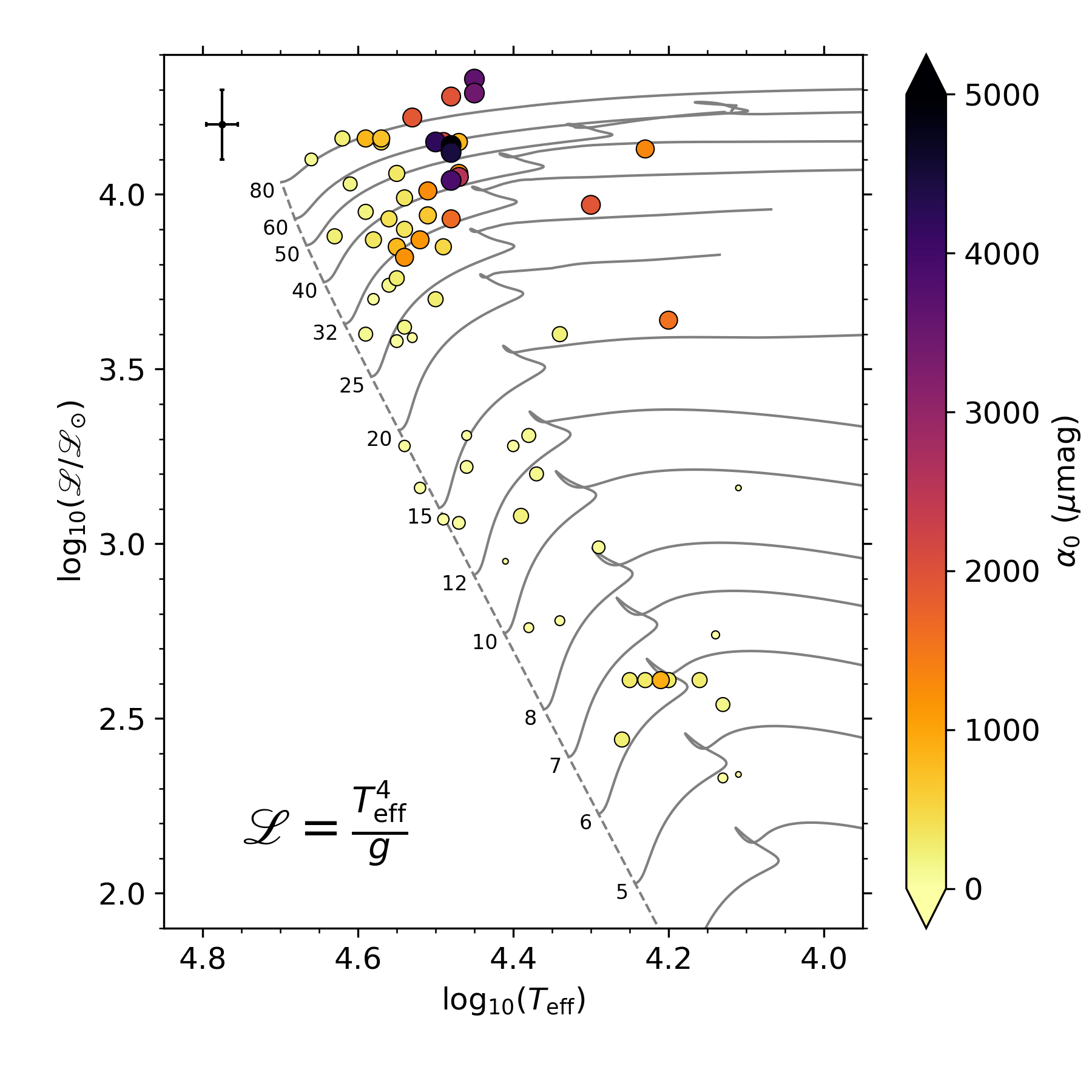} 
\includegraphics[width=0.49\textwidth]{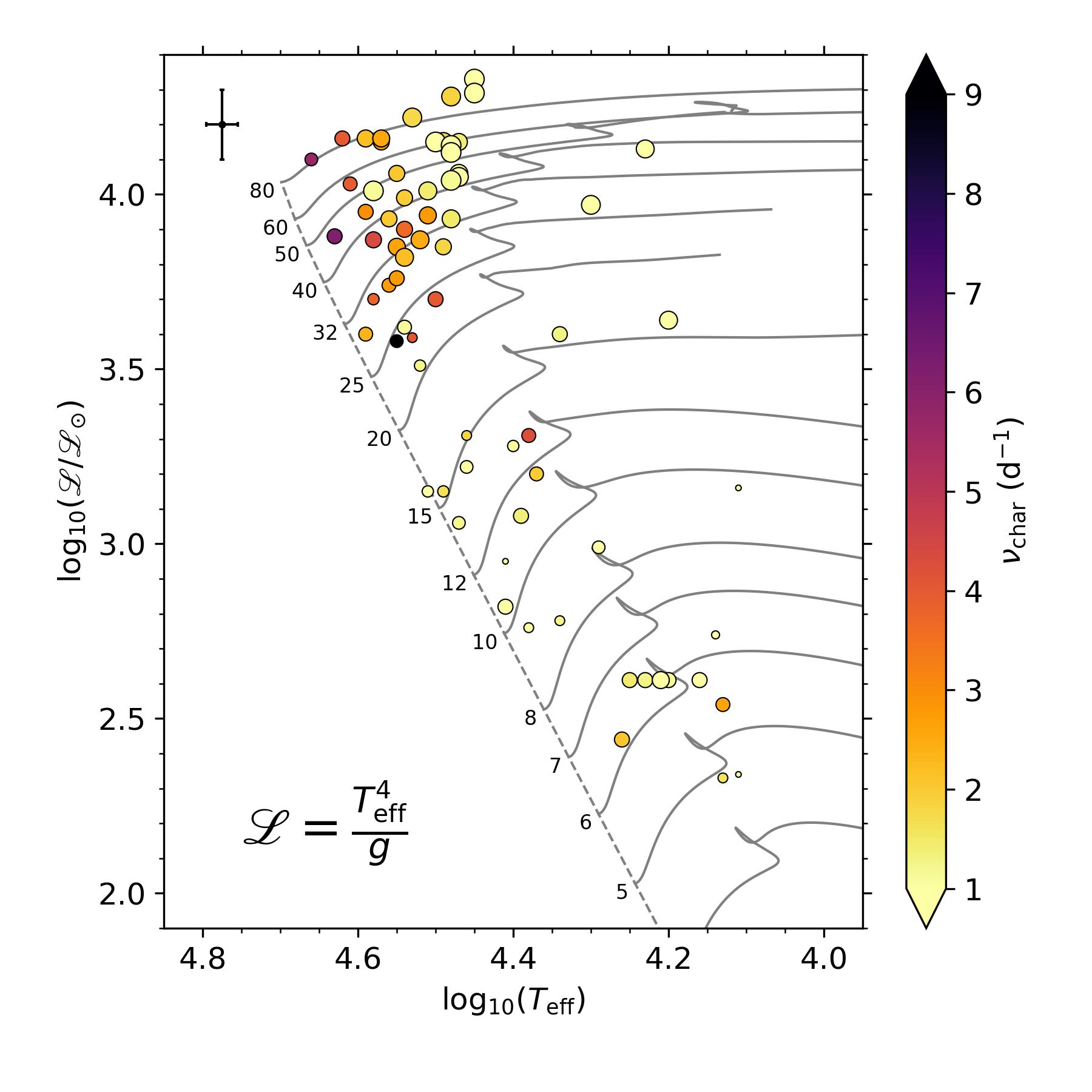} 
\caption{Spectroscopic HR~diagram of the massive star sample of \cite[Bowman et al. (2020)]{Bowman2020b} in which the morphology of SLF variability in the amplitude spectra of TESS light curves was quantified using Eqn.~(\ref{equation: red noise}). The left and right panels contain the same sample, but each star, shown as a circle, is colour-coded by its $\alpha_0$ and $\nu_{\rm char}$ parameter value, respectively. Figures adapted from \cite[Bowman et al. (2020)]{Bowman2020b}, their figure 2.}
\label{figure: HRD}
\end{center}
\end{figure}

Beyond the discovery and characterisation of SLF variability in massive stars, the study by \cite[Bowman et al. (2020)]{Bowman2020b} combined high-photometric and short-cadence TESS light curves of massive stars with high-resolution spectroscopy from the IACOB and OWN surveys \cite[(Sim{\'o}n-D{\'i}az et al. 2011; Burssens et al. 2020)]{Simon-Diaz2011, Burssens2020a}. For the first time, it was shown how the morphology of SLF variability, and specifically the characteristic amplitudes and frequencies, $\alpha_0$ and $\nu_{\rm char}$, directly probe a star's location in the Hertzsprung--Russell (HR) diagram. More massive and more evolved stars have larger amplitudes (i.e. $\alpha_0$) and lower frequencies (i.e. $\nu_{\rm char}$) in the morphology of their SLF variability, on average. Consequently, the characterisation of SLF variability provides constraints on a massive star's mass and age without the need for spectroscopic constraints. This is demonstrated in Fig.~\ref{figure: HRD}, in which the left and right panels denote each star's location in the spectroscopic HR~diagram colour-coded by $\alpha_0$ and $\nu_{\rm char}$, respectively.

An explanation for these trends is that more massive and more evolved stars have larger radii. This means that for larger radii, SLF variability exhibits longer periods, and vice versa. In terms of amplitudes, more evolved stars are observed to have larger amplitudes in their SLF variability, which means that the excitation mechanism(s) responsible exist from the zero-age main sequence (ZAMS) through to the terminal-age main sequence (TAMS), but are more efficient for more evolved stars. It is important to note that the sample of massive stars in Fig.~\ref{figure: HRD} is far from complete and suffers from biases. For example, the B-star regime is severely under sampled, as is the post-TAMS phase of evolution. Therefore, the observed trends in SLF variability with mass and age are only evident for main-sequence O~stars, and does include some astrophysical scatter.


\section{Correlation with Macroturbulence}

In addition to placing stars in the HR~diagram, \cite[Bowman et al. (2020)]{Bowman2020b} found a strong correlation between the characteristic amplitudes and frequencies of SLF variability and macroturbulence in massive stars. Macroturbulence is a non-rotational broadening mechanism needed to fit spectral lines of massive stars, which has both radial and tangential components and amplitudes of order 100~km\,s$^{-1}$ among O stars \cite[(Sim{\'o}n-D{\'i}az et al. 2014; Sim{\'o}n-D{\'i}az et al. 2017)]{Simon-Diaz2014, Simon-Diaz2017}. Macroturbulence is typically the dominant form of spectral line broadening in massive stars, so various studies have placed specific emphasis on investigating its physical cause (see e.g. \cite[Aerts et al. 2019; Grassitelli et al. 2015)]{Aerts2009, Grassitelli2015}.

\begin{figure}[b]
\begin{center}
\includegraphics[width=0.99\textwidth]{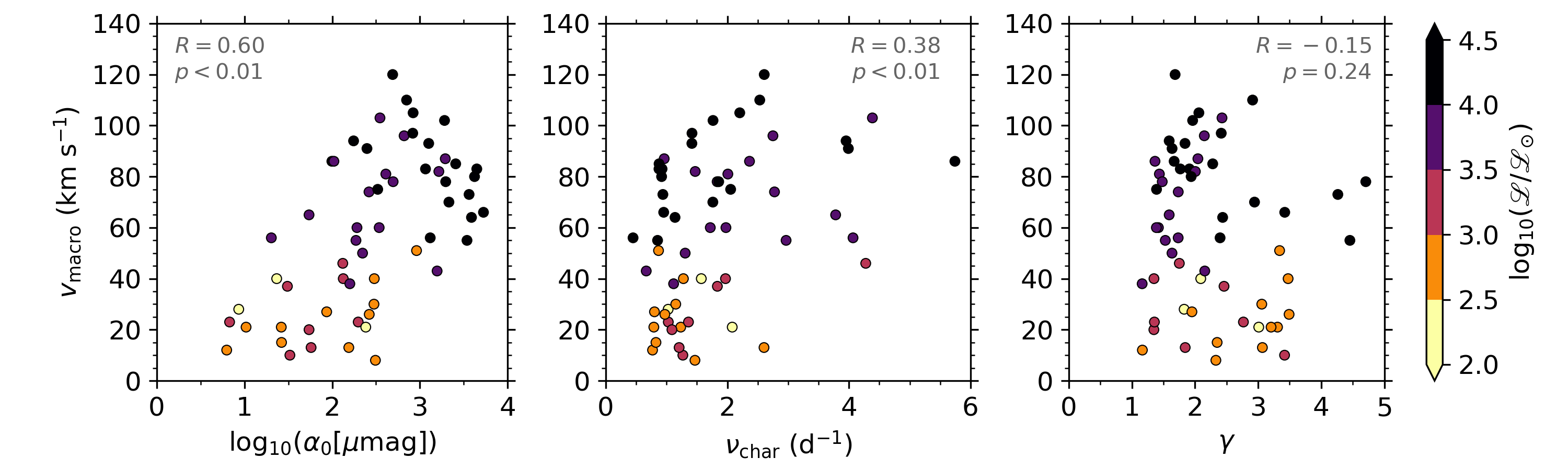} 
\caption{Pairwise correlation statistics between the amount of spectroscopic macroturbulence, $v_{\rm macro}$, and the morphology parameters of SLF variability determined by \cite[Bowman et al. (2020)]{Bowman2020b}. Figure adapted from \cite[Bowman et al. (2020)]{Bowman2020b}, their figure 3.}
\label{figure: vmacro}
\end{center}
\end{figure}

From their sample of massive stars, \cite[Bowman et al. (2020)]{Bowman2020b} found a statistically significant relationship between the amount of macroturbulence needed to fit the spectral line profiles and the characteristic amplitudes and frequencies of SLF variability in TESS photometry. Specifically, stars with larger macroturbulence have large values of $\alpha_0$ and smaller values of $\nu_{\rm char}$. This correlation is demonstrated in Fig.~\ref{figure: vmacro}, in which the correlation coefficient, $R$, and the resulting $p$-value from a hypothesis test is shown in the top corner of each panel. On the other hand, there is no significant correlation among the steepness, $\gamma$, of the SLF variability morphology, the amount of macroturbulence, nor the location of the star in the HR~diagram. This demonstrates that a typical $\gamma$ value of $1 \lesssim \gamma \lesssim 3$ exists for all massive stars, regardless of their mass, age or the physical mechanism(s) that cause SLF variability. 

Due to the limited sample size, it was not possible for \cite[Bowman et al. (2020)]{Bowman2020b} to investigate further correlations with, for example, rotation rate or metallicity. However, such an investigation with a much larger sample of massive stars will undoubtedly shine further light on the physical relationship between macroturbulence in spectroscopy and SLF variability in time-series photometry. For example, rotation is expected to significantly modify the structure of a star and also shift the frequencies of pulsation modes, which may explain the scatter in Figs.~\ref{figure: HRD} and \ref{figure: vmacro}.

It was first \cite[Lucy (1976)]{Lucy1976} who postulated that macroturbulence was caused by pulsations. Later, \cite[Aerts et al. (2009)]{Aerts2009} showed using detailed simulations that non-radial gravity modes provide the necessary predominantly tangential velocity field to reproduce spectral line profile broadening caused by macroturbulence. More recently, \cite[Aerts \& Rogers (2015)]{Aerts2015} extended this and demonstrated how an ensemble of stochastically excited gravity waves provides the necessary tangential velocity field to explain macroturbulence as a pulsational broadening mechanism. Together, this makes pulsations, specifically gravity waves, a plausible mechanism for macroturbulence and SLF variability in massive stars.


\section{Discussion and Conclusions}

The discovery of SLF variability in massive stars has inspired and energised various theoretical studies dedicated to providing possible explanations of this new type of observational signal. Motivated by there always being a convective core for main-sequence stars, \cite[Aerts \& Rogers (2015)]{Aerts2015} postulated that the SLF variability detected at the surface of three O stars was caused by stochastically excited gravity waves generated at the interface of the convective core and radiative envelope. This was based on the qualitatively similar amplitude spectra of these three O stars with those predicted from 2D hydrodynamical simulations \cite[(Rogers et al. 2013; Rogers 2015; Rogers \& McElwaine 2017)]{Rogers2013, Rogers2015, Rogers2017}. More recently, further 2D and 3D hydrodynamical simulations of stellar interiors using different numerical setups have shown that gravity waves excited at the boundaries of convective cores produce detectable surface velocity and flux perturbations \cite[(Edelmann et al. 2019; Horst et al. 2020)]{Edelmann2019, Horst2020}. Such predictions are indeed quantitatively comparable to the observed amplitude spectra of massive stars \cite[(Bowman et al. 2019b, 2020)]{Bowman2019b, Bowman2020b}.

On the other hand, the presence of a sub-surface convection zone and how its properties depend on a star's location in the HR~diagram has led other studies to postulate that SLF variability and/or macroturbulence originates much closer to the surface \cite[(Cantiello et al. 2009; Grassitelli et al. 2015; Cantiello et al. 2021)]{Cantiello2009, Cantiello2021}. Moreover, numerical calculations of wave transfer functions for gravity waves generated near the convective cores of main-sequence stars predict them to appear as a series of equally-spaced peaks in the amplitude spectrum, which correspond to the stellar eigenfrequencies, rather than as SLF variability \cite[(Lecoanet et al. 2019; Lecoanet et al. 2021)]{Lecoanet2019a, Lecoanet2021a}. However, the effects of non-linearity and rotation should be studied in detail in the context of wave propagation, because the Coriolis force is a significant restoring force for gravity waves in massive stars \cite[(Aerts et al. 2019)]{Aerts2019}. Furthermore, the dynamics of sub-surface convection zones in massive stars are expected to be quite turbulent. This is demonstrated by hydrodynamical simulations \cite[(Jiang et al. 2015, 2018; Schultz et al. 2022)]{Jiang_Y_2015, Jiang_Y_2018b, Schultz2022a}, which as a consequence also predict SLF variability in time-series photometry. However, the efficiency of sub-surface convection zones is strongly metallicity dependent, such that some stars may have no subsurface convection zones at all \cite[(Jermyn et al. 2022)]{Jermyn2022}. Finally, for more evolved stars near or beyond the TAMS, their radiatively driven and optically thick winds are predicted to produce SLF variability in time-series observations \cite[(e.g. Krti{\v c}ka \& Feldmeier 2018, 2021)]{Krticka2018, Krticka2021}. 

All of the aforementioned mechanisms that predict SLF variability in massive stars are not mutually exclusive. This means that the challenge ahead is in disentangling which mechanism dominates within specific parameter regimes in the HR~diagram. This remains an open question because the properties of (sub-surface) convection, winds, and the driving and propagation of gravity waves depend on the mass, radius, metallicity, and rotation rate of a massive star. What is clear is that any mechanism that aims to unanimously explain SLF variability in main-sequence stars must do so for stars that span a broad range of masses (i.e. $3 \lesssim M \lesssim 100$~M$_{\odot}$), ages from the ZAMS to the TAMS and beyond, rotation rates between zero and large fractions of the critical breakup velocity, and for both metal-poor (i.e. $Z \leq 0.002$) and metal-rich stars (i.e. $Z \geq 0.014$). 

Since the work by \cite[Bowman et al. (2019b) and Bowman et al. (2020)]{Bowman2019b, Bowman2020b}, much larger samples of massive stars with longer light curves are available thanks to the ongoing TESS mission. In the future, new ground-based spectroscopic monitoring campaigns and space missions, such as the CubeSpec mission \cite[(Bowman et al. 2022)]{Bowman2022}, will provide invaluable constraints on the time-dependent nature of macroturbulence and its link to pulsations in massive stars. From renewed theoretical work on understanding the interiors and atmospheres of massive stars, the constraints on the physical mechanisms that govern their structure and evolution are ever improving. Asteroseismology of SLF variability not only has the potential to provide independent masses and ages of massive stars, but also provide insight of their interior mixing and rotation properties to calibrate stellar structure and evolution theory.

\section*{Acknowledgements}

DMB gratefully acknowledges funding from the Research Foundation Flanders (FWO) by means of a senior postdoctoral fellowship (grant agreement No.~1286521N). DMB thanks the TESS science team for the excellent data, without which massive star asteroseismology and much of this work would not have been possible.





\begin{thebibliography}{}



\bibitem[Aerts et al. (2019)]{Aerts2019} {Aerts et al.} 2019, \textit{ARAA}, 57, 35-78

\bibitem[Aerts et al. (2009)]{Aerts2009} {Aerts et al.} 2009, \textit{A\&A}, 508, 409-419

\bibitem[Aerts \& Rogers (2015)]{Aerts2015} {Aerts \& Rogers} 2015, \textit{ApJL}, 806, L33

\bibitem[Aerts (2021)]{Aerts2021} {Aerts} 2021, \textit{Reviews of Modern Physics}, 93, 015001

\bibitem[Auvergne et al. (2009)]{Auvergne2009} {Auvergne et al.} 2009, \textit{A\&A}, 506, 411-424

\bibitem[Borucki et al. (2010)]{Borucki2010} {Borucki et al.} 2010, \textit{Science}, 327, 977

\bibitem[Bowman (2017)]{Bowman2017} {Bowman} 2017, \textit{Springer}, Amplitude modulation of pulsation modes in delta Scuti stars

\bibitem[Bowman et al. (2019)]{Bowman2019a} {Bowman et al.} 2019, \textit{A\&A}, 621, A135

\bibitem[Bowman et al. (2019)]{Bowman2019b} {Bowman et al.} 2019, \textit{Nature Astronomy}, 3, 760-765

\bibitem[Bowman et al. (2020)]{Bowman2020b} {Bowman et al.} 2020, \textit{A\&A}, 640, A36

\bibitem[Bowman et al. (2022)]{Bowman2022} {Bowman et al.} 2022, \textit{A\&A}, 658, A96

\bibitem[Bowman (2020)]{Bowman2020c} {Bowman} 2020, \textit{Fron. Astron. Space Sci.}, 7, 70

\bibitem[Burssens et al. (2020)]{Bowman2020} {Burssens et al.} 2020, \textit{A\&A}, 639, A81

\bibitem[Cantiello et al. (2009)]{Cantiello2009} {Cantiello et al.} 2009, \textit{A\&A}, 499, 279-290

\bibitem[Cantiello et al. (2021)]{Cantiello2021} {Cantiello et al.} 2021, \textit{ApJ}, 915, 112

\bibitem[Dorn-Wallenstein et al. (2020)]{Dorn-Wallenstein2020} {Dorn-Wallenstein et al.} 2020, \textit{ApJ}, 878, 155

\bibitem[Edelmann et al. (2019)]{Edelmann2019} {Edelmann et al.} 2019, \textit{ApJ}, 876, 4

\bibitem[Elliott et al. (2022)]{Elliott2022} {Elliott et al.} 2022, \textit{MNRAS}, 509, 4246-4255

\bibitem[Grassitelli et al. (2015)]{Grassitelli2015} {Grassitelli et al.} 2015, \textit{ApJL}, 808, L31

\bibitem[Horst et al. (2020)]{Horst2020} {Horst et al.} 2020, \textit{A\&A}, 641, A18

\bibitem[Howell et al. (2014)]{Howell2014} {Howell et al.} 2014, \textit{PASP}, 126, 398-408

\bibitem[Jermyn et al. (2022)]{Jermyn2022} {Jermyn et al.} 2022, \textit{ApJ}, 926, 221

\bibitem[Jiang et al. (2015)]{Jiang2015} {Jiang et al.} 2015, \textit{ApJ}, 813, 74

\bibitem[Jiang et al. (2018)]{Jiang2018} {Jiang et al.} 2018, \textit{Nat}, 561, 498-501

\bibitem[Krti{\v c}ka \& Feldmeier (2018)]{Krticka2018} {Krti{\v c}ka \& Feldmeier et al.} 2018, \textit{A\&A}, 617, A121

\bibitem[Krti{\v c}ka \& Feldmeier (2021)]{Krticka2021} {Krti{\v c}ka \& Feldmeier et al.} 2021, \textit{A\&A}, 648, A79

\bibitem[Lecoanet et al. (2019)]{Lecoanet2019} {Lecoanet et al.} 2019 \textit{ApJL}, 886, L15

\bibitem[Lecoanet et al. (2021)]{Lecoanet2021} {Lecoanet et al.} 2021 \textit{MNRAS}, 508, 132-143

\bibitem[Lenoir-Craig et al. (2022)]{Lenoir-Craig2022} {Lenoir-Craig et al.} 2022 \textit{ApJ}, 925, 79

\bibitem[Lucy (1976)]{Lucy1976} {Lucy} 1976 \textit{ApJ}, 206, 499

\bibitem[Naz{\'e} et al. (2021)]{Naze2021} {Naz{\'e} et al.} 2021, \textit{MNRAS}, 502, 5038-5048

\bibitem[Ricker et al. (2015)]{Ricker2015} {Ricker et al.} 2015, \textit{J. Astron. Telesc., Instrum., Sys.}, 1, 014003

\bibitem[Rogers et al. (2013)]{Rogers2013} {Rogers et al.} 2013, \textit{ApJ}, 772, 21

\bibitem[Rogers (2015)]{Rogers2015} {Rogers} 2015, \textit{ApJL}, 815, L30

\bibitem[Rogers \& McElwaine (2017)]{Rogers2017} {Rogers \& McElwaine} 2017, \textit{ApJL}, 848, L1

\bibitem[Schultz et al. (2022)]{Schultz2022} {Schultz et al.} 2022, \textit{ApJL}, 924, L11

\bibitem[Sim{\'o}n-D{\'{\i}}az et al. (2011)]{Simon-Diaz2011} {Sim{\'o}n-D{\'{\i}}az et al.} 2011, \textit{Bulletin de la Societe Royale des Sciences de Liege}, 80, 514-518

\bibitem[Sim{\'o}n-D{\'{\i}}az et al. (2014)]{Simon-Diaz2014} {Sim{\'o}n-D{\'{\i}}az et al.} 2014, \textit{A\&A}, 562, A135

\bibitem[Sim{\'o}n-D{\'{\i}}az et al. (2017)]{Simon-Diaz2017} {Sim{\'o}n-D{\'{\i}}az et al.} 2017, \textit{A\&A}, 597, A22




\end{thebibliography}
\end{document}